\documentclass[preprintnumbers,amsmath,amssymb,floatfix,11pt,prd,onecolumn,superscriptaddress,nofootinbib]{revtex4}
\usepackage{graphicx}
\usepackage{epsfig}
\usepackage{bm}
\usepackage{amsfonts}

\begin{document}

\title{ Noether gauge symmetry for Bianchi type I model in $f(T)$ gravity}

\author{\textbf{ Adnan Aslam}}
\affiliation{Center for Advanced Mathematics and Physics (CAMP),\\
National University of Sciences and Technology (NUST), H-12,
Islamabad, Pakistan}

\author{\textbf{ Mubasher Jamil}} \email{mjamil@camp.nust.edu.pk}
\affiliation{Center for Advanced Mathematics and Physics (CAMP),\\
National University of Sciences and Technology (NUST), H-12,
Islamabad, Pakistan}
\affiliation{Eurasian International Center
for Theoretical Physics,  L.N. Gumilyov Eurasian National University, Astana
010008, Kazakhstan}

\author{\textbf{ Ratbay Myrzakulov}}
\email{rmyrzakulov@gmail.com}\affiliation{Eurasian International Center
for Theoretical Physics,  L.N. Gumilyov Eurasian National University, Astana
010008, Kazakhstan}

\begin{abstract}
\textbf{Abstract:} In this Letter, we  have presented the Noether symmetries of a class of Bianchi type I anisotropic model in the context of the $f(T)$ gravity. By solving the system of equations obtained from the Noether symmetry condition, we obtain the form of the $f(T)$ as a Teleparallel form. This analysis shows that the teleparallel gravity has the maximum number of Noether symmetries. We derive the symmetry generators and show that there are five kinds of the symmetries including time and scale invariance under metric coefficients. We classified the symmetries and further we obtain the corresponding invariants.
\end{abstract}
\maketitle

\newpage

\section{Introduction}

In the last decade, one of the big challenge for researches in physicist
community is explanation of the essence and mechanism of the acceleration of our universe
\cite{expansion} in this era of the universe, which has confirmed by some observation data
such as supernova type Ia \cite{Supernova}, baryon acoustic
oscillations \cite{oscillations}, weak lensing \cite{lensing} and
large scale structure \cite{LSS}. Finding the phenomenological explanation of cosmic acceleration has been
one of the main problems of cosmology and high energy theoretical physics \cite{smolin}. Reviews of some recent and old attempts to resolve the issue of dark energy and related problems can be found in \cite{reviews}.

In order to explain the current accelerated expansion without
introducing dark energy, one may use a simple generalized version
of the so-called teleparallel gravity \cite{Albert}, namely $f(T)$
theory. Although teleparallel gravity is not an alternative to
general relativity (they are dynamically equivalent), but its
different formulation allows one to say: gravity is not due to
curvature, but to torsion. In other word, using curvature-less
Weitzenbck connection instead of torsion-less Levi-Civita
connection in standard general relativity leads to, subsequently,
replacing curvature by torsion. We should note that one of the main requirement of $f(T)$ gravity
is that there exist a class of spin-less connection frames where
its torsion does not vanish \cite{Barrow}. Considering the above
$f(T)$ theory leads to interesting cosmological behavior and its
various aspects have been examined in the literature \cite{F(T)}.

On the other side, we could not ignore the important role of
continuous symmetry in the mathematical physics. In particular,
the well-known Noether's symmetry theorem is a practical tool in
theoretical physics which states that any differentiable symmetry
of an action of a physical system leads to a corresponding
conserved quantity, which so called Noether charge \cite{Noether}.
In the literature, applications of the Noether symmetry in
generalized theories of gravity have been studied (see
\cite{ApplNoether1} and references therein).
 In particular, Wei et al calculated Noether symmetries of $f(T)$ cosmology containing
  matter and found a power-law solution $f(T)\sim\mu T^n$
  \cite{ApplNoether1}. Further they showed that if $n > 3/2$ the expansion of our universe can be
    accelerated without invoking dark energy. We reconsider their model in the
    frame of the anisotropic Bianchi type I models, and find that the
     corresponding symmetries. Noether gauge symmetry is also
     applied to $f(T)$ gravity minimally coupled with a canonical
     scalar field to determine the unknown functions of the theory:
     $f(T), V(\phi), W(\phi)$ and it is shown that the behavior of
     the hubble parameter in the model closely matches to the
     $\Lambda$CDM  model \cite{A1}. A detail discussion on the
     Noether symmetries and conserved quantities is given in
     \cite{A2}. An investigation of the Noether symmetries of $f(T)$
     cosmology involving matter and dark energy, where dark energy
     is  represented by a canonical scalar field with potential and it is shown that
     $f(T)\sim T^{\frac{3}{4}}$ and $V(\phi)\sim \phi^{2}$
     \cite{M1}. To explain the accelerated expansion of the universe
     $f(T)$ gravity is considered with cold dark matter and Noether
     symmetries are utilized to study the cosmological consequences
     \cite{HMS}.
  Recently we applied the Noether-Gauge symmetry in Bianchi type I models in some generalized scalar field models \cite{epjc}. Further, there are some exact solutions for Bianchi type I models in $f(T)$ gravity \cite{sharif}. In this paper we want to find the symmetries of a Bianchi type I spacetime filled with perfect fluid with barotropic EoS using the Noether gauge symmetry approach.

The plan of our Letter is as follows: In section-II, we review the basics of the $f(T)$ gravity.   In section-III, we write down the Lagrangian of our model and related dynamical equations. In section-IV we present the Noether symmetries and invariants of our model.  Finally we conclude the results in section-V.

\section{Basics of $f(T)$ gravity}

General relativity is a gauge theory of the gravitational field. It is based on the equivalence principle. However  it is not neccesary to working with Riemannian manifolds. There are some extended theories as Riemann-Cartan, in them the geometrical structure of the theory is not the metricity. In these extensions, there are more than one dynamical quantity (metric). For example this theory may be constructed from the  metric, non-metricity and torsion  \cite{smalley}. Ignoring from the non-metricity of the theory, we can leave the Riemannian manifold and going to the Weitznbock spacetime, with torsion and zero local Riemann tensor. One sample of such theories  is called teleparallel gravity in which we are working in a non-Riemannian manifold. The dynamics of the metric determined using the scalar torsion $T$. The fundamental quantities in teleparallel theory is the vierbein (tetrad) basis  $e^{i}_{\mu}$. This basis is an orthogonal, coordinate free basis, defined by the following equation
\begin{eqnarray}\nonumber
g_{\mu\nu}=\eta_{ij}e_{\mu}^{i}e_{\nu}^j .
\end{eqnarray}
This tetrade basis must be orthogonormal and $\eta_{ij}$ is the Minkowski metric. It means that $e^{i}_{\mu}e^{\mu}_{ j}=\delta^i_{j}$.
There is a simple extension of the teleparallel gravity, called  $f(T)$ gravity, where $f$ is an arbitrary function of the torsion $T$. One suitable form of  action for $f(T)$ gravity in  Weitzenbock
manifold is \cite{saridakis}
\begin{eqnarray}\nonumber
\mathcal{S}=\frac{1}{2\kappa^2}\int d^4x \sqrt{e}(T+f(T)+\mathcal{L}_m).
\end{eqnarray}
Here $e=det(e^{i}_{\mu})$, $\kappa^2=8\pi G$. The dynamical quantity of the model is the scalar torsion $T$ and $L_m$ is the matter Lagrangian.
The field equation can be derived from the action by varying the action with respect to $e^{i}_{\mu}$, which reads
\begin{eqnarray}\nonumber
&&e^{-1}\partial_{\mu}(e S^{\:\:\:\mu
\nu}_{i})(1+f_T)-e_i^{\:\lambda}T_{\:\:\:\mu
\lambda}^{\rho}S^{\:\:\:\nu \mu}_{\rho}f_T +S^{\:\:\:\mu
\nu}_{i}\partial_{\mu}(T)f_{TT}-\frac{1}{4}e_{\:i}^{\nu}
(1+f(T))=4 \pi G e_i^{\:\rho}T_{\rho}^{\:\:\nu},
\end{eqnarray}
where $T_{\rho}^{\:\:\nu}$ is the energy-momentum tensor for matter sector of the Lagrangian $\mathcal{L}_m$, defined by $$T_{\mu\nu}=-\frac{2}{\sqrt{-g}}\frac{\delta\int\Big(\sqrt{-g}\mathcal{L}_m d^4x\Big)}{\delta g^{\mu\nu}}.$$ Here $T$ is defined by
\begin{equation}\nonumber
T=S^{\:\:\:\mu \nu}_{\rho} T_{\:\:\:\mu \nu}^{\rho},
\end{equation}
where
$$
T_{\:\:\:\mu \nu}^{\rho}=e_i^{\rho}(\partial_{\mu}
e^i_{\nu}-\partial_{\nu} e^i_{\mu}),
$$
$$
S^{\:\:\:\mu \nu}_{\rho}=\frac{1}{2}(K^{\mu
\nu}_{\:\:\:\:\:\rho}+\delta^{\mu}_{\rho} T^{\theta
\nu}_{\:\:\:\theta}-\delta^{\nu}_{\rho} T^{\theta
\mu}_{\:\:\:\theta}),
$$
and the contorsion tensor reads $K^{\mu \nu}_{\:\:\:\:\:\rho}$  as
$$
K^{\mu \nu}_{\:\:\:\:\:\rho}=-\frac{1}{2}(T^{\mu
\nu}_{\:\:\:\:\:\rho}-T^{\nu \mu}_{\:\:\:\:\:\rho}-T^{\:\:\:\mu
\nu}_{\rho}).
$$
It is staighforward to show that this equation
 of motion reduces to the Einstein gravity when $f(T)=0$. Indeed, it's the equivalency between the teleparallel theory and Einstein gravity \cite{T}. The theory has been found to address the issue of cosmic acceleration in the early and late evolution of universe \cite{godel} but this crucially depends on the choice of suitable $f(T)$, for instance exponential form containing $T$ can not lead to phantom crossing \cite{wuyu}.
 Reconstruction of $f(T)$ models has been reported in \cite{setare} while thermodynamics
of $f(T)$ cosmology including generalized second law of thermodynamics is recently investigated \cite{gsl}.

\section{The model }

 We start from the standard gravitational
action (chosen units are $c=8\pi G=1$) \cite{sharif}
\begin{equation}
\mathcal{S}=\int d^{4}x\sqrt{-g}\Big[T+f(T)-(1+f_T(T))\Big[T+2\Big(\frac{\dot{A}\dot{B}}{AB}+\frac{\dot{A}\dot{C}}{AC}+\frac{\dot{C}\dot{B}}{CB}\Big)\Big]+\mathcal{L}_m\Big],
\end{equation}
 We
consider the Bianchi - I  metric
\begin{equation}\label{1}
ds^2=-dt^2+A^2dx^2+B^2dy^2+C^2dy^2.
\end{equation}
Here the metric potentials $A$, $B$ and $C$ are functions of $t$ alone.  We as assume that the spacetime filling with a perfect fluid.
The Lagrangian of Bianchi type-I  model can be written as
\begin{equation}\label{L}
\mathcal{L}(A,B,C,T,\dot{A},\dot{B},\dot{C})=ABC\Big[T+f(T)-(1+f_T)\Big[T+2\Big(\frac{\dot{A}\dot{B}}{AB}+\frac{\dot{A}\dot{C}}{AC}+\frac{\dot{C}\dot{B}}{CB}\Big)\Big]+\mathcal{L}_m\Big].
\end{equation}
Note that the Lagrangian does not depend on $\dot T$.   We must emphasize here that when the global geometry of spacetime remains homogeneous, for both isotropic and anisotropic cases, there exists an isothermal process for tending the whole system to the maximum entropy. Via such isothermal process, it is possible to define a global (not local) average pressure. In an expanding universe, we define an everage Hubble parameter  by $H=\frac{1}{3}\Sigma_{i} H_i$, where $H_i=\frac{\dot A_i}{A_i}$. Such average Hubble parameter leads to the effective global pressure of the anisotropic spacetime \cite{p}.

The equations of motion corresponding to Lagrangian (\ref{L}) are
\begin{eqnarray}
f_{TT}\Big[T+2\Big(\frac{\dot{A}\dot{B}}{AB}+\frac{\dot{A}\dot{C}}{AC}+\frac{\dot{C}\dot{B}}{CB}\Big)\Big]=0,\\
2\dot Tf_{TT}\Big( \frac{\dot B}{B}+\frac{\dot C}{C} \Big)+2(1+f_T)\Big( \frac{\ddot B}{B}+2\frac{\dot B}{B}\frac{\dot C}{C}+\frac{\ddot C}{C} \Big)=2(1+f_T)\frac{\dot B}{B}\frac{\dot C}{C}-(f-Tf_T)-p_1,\\
2\dot Tf_{TT}\Big( \frac{\dot A}{A}+\frac{\dot C}{C} \Big)+2(1+f_T)\Big( \frac{\ddot A}{A}+2\frac{\dot A}{A}\frac{\dot C}{C}+\frac{\ddot C}{C} \Big)=2(1+f_T)\frac{\dot A}{A}\frac{\dot C}{C}-(f-Tf_T)-p_2,\\
2\dot Tf_{TT}\Big( \frac{\dot B}{B}+\frac{\dot A}{A} \Big)+2(1+f_T)\Big( \frac{\ddot B}{B}+2\frac{\dot B}{B}\frac{\dot A}{A}+\frac{\ddot A}{A} \Big)=2(1+f_T)\frac{\dot B}{B}\frac{\dot A}{A}-(f-Tf_T)-p_3,
\end{eqnarray}
From equation (4), there are two possibilities: (1) $f_{TT}=0$, which indicates teleparallel gravity, and we will be back to this case later, (2) other possibility is $$T=-2\Big(\frac{\dot{A}\dot{B}}{AB}+\frac{\dot{A}\dot{C}}{AC}+\frac{\dot{C}\dot{B}}{CB}\Big),$$ which is the definition of scalar torsion for Bianch I model. Further it is easy to show that by applying $e^i_\mu=diag(1,A,B,C)$, we can obtain the definition of torsion scalar. By substituting $f=0$ and $A=B=C=a$, we obtain the usual second Friedmann equation. We can redefine the three anisotropic pressure as the components of energy-momentum tensor  $T_{\mu\nu}=diag(\rho,-p_1,-p_2,-p_3).$

Adding Eqs. (5),(6) and (7) yields
$$ 12\dot THf_{TT}+2(1+f_T)[3(\dot H+\sum H_i^2)]-T(1+f_T)=-3(f-Tf_T)-P,  $$
$$ -8\dot HTf_{TT}-(4\dot H-T)-(4\dot H-2T)f_T-f=\rho. $$

The first Friedmann equation in this model is obtained by the Hamiltonian constraint equation \cite{sharif}
\begin{equation}
-T-2Tf_T+f=\rho.
\end{equation}
In this paper, we will be working only with $p_1=p_2=p_3$ which comes from the definition of pressure $P$.

\section{Noether gauge symmetry of the model}

In this section, we carry out the Noether gauge symmetry equations for the matter Lagrangian $\mathcal{L}_m=P=P_0(ABC)^{1/(1+w)}$.

A vector field
\begin{eqnarray}
\mathbf{X}&=&\xi(t,A,B,C,T)\frac{\partial }{\partial
t}+\eta^{1}(t,T,A,B,C)\frac{\partial }{\partial
T}+\eta^{2}(t,T,A,B,C)\frac{\partial }{\partial
A}\nonumber\\&&+\eta^{3}(t,T,A,B,C)\frac{\partial }{\partial
B}+\eta^{4}(t,T,A,B,C,T)\frac{\partial }{\partial C},
\end{eqnarray}
is a Noether gauge symmetry of the Lagrangian \cite{bb1}, if there
exists a vector valued gauge function $\mathcal{G}\in{\mathcal{U}},$ where
$\mathcal{U}$ is the space of differential functions, such that
\begin{equation}
\mathbf{X^{[1]}}\mathcal{L}+\mathcal{L}(D_{t}\xi)=D_{t}\mathcal{G},\label{NC}
\end{equation}
where $\mathcal{L}$ is the the Lagrangian (\ref{L}), $D_{t}\xi$
represents the total derivative of $\xi$ and the coefficients $\xi,~
\eta^{i}, ~ (i=1,2,3,4)$ are determined from the Noether symmetry
conditions. Now if one want to apply the generator $\mathbf{X}$ to
the Lagrangian or differential equation consisting of the
independent variables, the dependent variables and the derivatives
of the dependent variables with respect to the independent variable
up to the order $n$. Then one have to prolong (or to extend) the
generator $\mathbf{X}$ up to $n$-th derivative, called the $n$-th
order prolongation, so that one can have the action of the generator
on all the derivatives. The space, whose coordinates represent the
independent variables, the dependent variables and the derivatives
of the dependent variables up to order $n$ is called the $n$-th
order jet space of the underlying space consisting of only the
independent and dependent variables. Here the Lagrangian contains
only the first order derivative of the dependent variable along with
the independent variables and the dependent variables. So we need
the first order prolongation of the above symmetry in the
first-order jet space comprising of all derivatives, which is given
by
\begin{equation*}
\mathbf{X}^{[1]} = \textbf{X}+{\eta}^{1}_{(t)} \frac{\partial}{\partial
\dot{T}} + {\eta} ^{2}_{(t)} \frac{\partial}{\partial \dot{A}} +
{\eta}^{3}_{(t)} \frac{\partial}{\partial \dot{B}}+ {\eta}^{4}_{(t)}
\frac{\partial}{\partial \dot{C}},
\end{equation*}
where
\begin{align}
{\eta}^{1}_{(t)}&=D_{t}\eta^{1}-\dot{T}D_{t}\xi, ~~~  {\eta}^{2}_{(t)}=D_{t}\eta^{2}-\dot{A}D_{t}\xi,\nonumber\\
{\eta}^{3}_{(t)}&=D_{t}\eta^{3}-\dot{B}D_{t}\xi,~~~{\eta}^{4}_{(t)}=D_{t}\eta^{4}-\dot{C}D_{t}\xi,.
\end{align}
and $D_{t}$ is the total derivative operator
\begin{equation}
D_{t}=\frac{\partial }{\partial t}+\dot{T}\frac{\partial }{\partial
T}+\dot{A}\frac{\partial }{\partial A}+\dot{B}\frac{\partial
}{\partial B}+\dot{C}\frac{\partial }{\partial C}.
\end{equation}
If $\mathbf{X}$ is the Noether symmetry corresponding to the
Lagrangian $\mathcal{L}(t,T,A,B,C,\dot{T},\dot{A},\dot{B},\dot{C})$, then
\begin{equation}
\mathbf{I}=\xi \mathcal{L}+(\eta^{1}-\xi\dot{T})\frac{\partial \mathcal{L}}{\partial
\dot{T}}+(\eta^{2}-\xi\dot{A})\frac{\partial \mathcal{L}}{\partial
\dot{A}}+(\eta^{3}-\xi\dot{B})\frac{\partial \mathcal{L}}{\partial
\dot{B}}+(\eta^{4}-\xi\dot{C})\frac{\partial \mathcal{L}}{\partial \dot{C}}-\mathcal{G},
\label{firstint}
\end{equation}
is a first integral or an invariant or a conserved quantity
associated with $\mathbf{X}$. The Noether symmetry condition
(\ref{NC}) yields the following system of linear partial differential equations (PDEs)
\begin{align}
&\mathcal{G}_{T}=0,~~\xi_{T}=0,~~\xi_{A}=0,~~\xi_{B}=0,~~\xi_{C}=0,\label{DE1}\\
&C\eta^{3}_{A}+B\eta^{4}_{A}=0,~~C\eta^{3}_{T}+B\eta^{4}_{T}=0,~~C\eta^{2}_{B}+A\eta^{4}_{B}=0,\\
&2(1+f_T)(C\eta^{3}_{t}+B\eta^{4}_{t})+\mathcal{G}_{A}=0,~~B\eta^{2}_{C}+A\eta^{3}_{C}=0,\\
&2(1+f_T)(B\eta^{2}_{t}+A\eta^{3}_{t})+\mathcal{G}_{C}=0,~~B\eta^{2}_{T}+A\eta^{3}_{T}=0,\\
&2(1+f_T)(C\eta^{2}_{t}+A\eta^{4}_{t})+\mathcal{G}_{B}=0,~~C\eta^{2}_{T}+A\eta^{4}_{T}=0,\\
&Af_{TT}\eta^{1}+(1+f_T)(\eta^{2}+C\eta^{2}_{C}+B\eta^{2}_{B}+A\eta^{3}_{B}+A\eta^{4}_{C}-A\xi_{t})=0,\\
&Bf_{TT}\eta^{1}+(1+f_T)(\eta^{3}+B\eta^{2}_{A}+C\eta^{3}_{C}+A\eta^{3}_{A}+B\eta^{4}_{C}-B\xi_{t})=0,\\
&Cf_{TT}\eta^{1}+(1+f_T)(\eta^{4}+C\eta^{2}_{A}+C\eta^{3}_{B}+B\eta^{4}_{B}+A\eta^{4}_{A}-C\xi_{t})=0,\\
&-ABCf_{TT}\eta^{1}+[BC(f-Tf_T)-wP_{0}BC(ABC)^{-1-w}]\eta^{2}\nonumber\\
&+[AC(f-Tf_T)-wP_{0}AC(ABC)^{-1-w}]\eta^{3}+[AB(f-Tf_T)-wP_{0}AB(ABC)^{-1-w}]\eta^{4}\nonumber\\
&+[ABC(f-Tf_T)+P_{0}(ABC)^{-w}]\xi_{t}=\mathcal{G}_{t}.\label{DEend}
\end{align}
The above system of linear PDEs yields four Noether symmetries
comprising of translation, rotation and other symmetries. This
system also specifies the form of the arbitrary function $f(T)$ as
follows,
\begin{align}
f(T)&=m T^{n},\\
\textbf{X}_{1}&=\frac{\partial}{\partial t},\\
\textbf{X}_{2}&=A\frac{\partial}{\partial A}-C\frac{\partial}{\partial C},\\
\textbf{X}_{3}&=B\frac{\partial}{\partial B}-C\frac{\partial}{\partial C},\\
\textbf{X}_{4}&=A\ln\Big(\frac{B}{C}\Big)\frac{\partial}{\partial
A}+B\ln\Big(\frac{C}{A}\Big)\frac{\partial}{\partial
B}+C\ln\Big(\frac{A}{B}\Big)\frac{\partial}{\partial C},
\end{align}
where $m$ and $n$ are arbitrary constants. $\textbf{X}_{1}$ to
$\textbf{X}_{4}$ correspond to either zero or constant gauge
function. For $n=1$, $f(T)$ becomes linear and the above system of
linear PDEs gives extra symmetries, along with the above mentioned
four symmetries, which are
\begin{align}
\textbf{X}_{5}&=t\frac{\partial}{\partial
t}+C\frac{\partial}{\partial C},\\
\textbf{X}&=g(t,T,A,B,C)\frac{\partial}{\partial T},
\end{align}
where $g(t,T,A,B,C)$ is an arbitrary function of it arguments. This
shows that we have the maximum number of Noether symmetries for the
linear case.

The invariants or conserved quantities corresponding to symmetry
generators are
\begin{eqnarray}
\mathbf{I}_{1}&=&2(1+m)(\dot{A}B\dot{C}+\dot{A}\dot{B}C+A\dot{B}\dot{C})+P_0(ABC)^{-w},\label{inter1}\\
\mathbf{I}_{2}&=&2(1+m)(\dot{A}BC-AB\dot{C}),\\
\mathbf{I}_{3}&=&2(1+m)(A\dot{B}C-AB\dot{C}),\\
\mathbf{I}_{4}&=&2(1+m)\Big[\dot{A}BC\ln\Big(\frac{B}{C}\Big)+A\dot{B}C\ln\Big(\frac{C}{A}\Big)
+AB\dot{C}\ln\Big(\frac{A}{B}\Big)\Big],\\
\mathbf{I}_{5}&=&2(1+m)t(\dot{A}B\dot{C}+\dot{A}\dot{B}C+A\dot{B}\dot{C}
)+tP_{0}(ABC)^{-w}-2(1+m)(\dot{A}BC+A\dot{B}C).\label{inter2}
\end{eqnarray}
From the previous remarks on the symmetries properties of the model,
as $\mathbf{X}_{1}=\frac{\partial}{\partial t}$ so $\mathbf{I}_{1}$
is the total energy or Hamiltonian of the system ,
 $\mathbf{I}_{2}$ and $\mathbf{I}_{3}$ characterize some conserved rotational momenta of the system,
  $\mathbf{I}_{4}$ and $\mathbf{I}_{5}$ is the rescaling momentum
  invariance.\\
For $A=B=C$ the invariants $\mathbf{I}_{1}$ to $\mathbf{I}_{5}$
becomes
\begin{eqnarray}
\mathbf{I}_{1}&=&6(1+m)(A\dot{A}^{2})+P_0(A)^{-3w},\label{inter1}\\
\mathbf{I}_{2}&=&0,\\
\mathbf{I}_{3}&=&0,\\
\mathbf{I}_{4}&=&0,\\
\mathbf{I}_{5}&=&6(1+m)t(A\dot{A}^{2})+tP_0(A)^{-3w}-4(1+m)(A^{2}\dot{A}).\label{inter2.1}
\end{eqnarray}
  The Equations of motion for the Lagrange (\ref{L}) corresponding to the case $f(T)=m
  T$ are
\begin{eqnarray}
T+2\Big(\frac{\dot{A}\dot{B}}{AB}+\frac{\dot{A}\dot{C}}{AC}+\frac{\dot{C}\dot{B}}{CB}\Big)=0,\label{EL
linear1}\\
2(1+m)\Big( \frac{\ddot B}{B}+\frac{\dot B}{B}\frac{\dot C}{C}+\frac{\ddot C}{C} \Big)-wP_{0}(ABC)^{-(w+1)}=0,\\
2(1+m)\Big( \frac{\ddot A}{A}+\frac{\dot A}{A}\frac{\dot C}{C}+\frac{\ddot C}{C} \Big)-wP_{0}(ABC)^{-(w+1)}=0,\\
2(1+m)\Big( \frac{\ddot B}{B}+2\frac{\dot B}{B}\frac{\dot
A}{A}+\frac{\ddot A}{A} \Big)-wP_{0}(ABC)^{-(w+1)}=0.\label{EL
linear2}
\end{eqnarray}
Now the total energy $E_{L}$ corresponding to $(0,0)$-Einstein
equation is
\begin{equation*}
E_{L}=\dot{T}\frac{\partial \mathcal{L}}{\partial
\dot{T}}+\dot{A}\frac{\partial \mathcal{L}}{\partial
\dot{A}}+\xi\dot{B}\frac{\partial \mathcal{L}}{\partial
\dot{B}}+\dot{C}\frac{\partial \mathcal{L}}{\partial
\dot{C}}-\mathcal{L},
\end{equation*}
which is equal to $-\mathbf{I}_{1}$, so $\mathbf{I}_{1}=0$.
$\mathbf{I}_{2}$ and $\mathbf{I}_{3}$ are conserved quantities so
these must be constants say $\mathbf{I}_{1}=a$ and
$\mathbf{I}_{2}=b$,where $a$ and $b$ are constants.\\
 These are the constraints which the equations of motion (\ref{EL
 linear1})-(\ref{EL linear2}) must satisfy. These are sufficient to
 find the solution of the equations of motion. Under these
 constraints we have the solution of the equations of motion (\ref{EL
 linear1})-(\ref{EL linear2}) for $\mathcal{L}_{m}=0$ as follows
\begin{eqnarray}
A(t)&=&\Big[\frac{1}{(-b+a)a}\big[(c_{3}+c_{2}t)(ba-b^{2}-a^{2}\mp(-b+2a)\sqrt{a^{2}-ba+b^{2}})
\big]\Big]^{\big(\frac{a}{b}\big)^{\pm 1}},\label{a1}\\
B(t)&=&c_{1}\exp{\int\dot{A}({\dot{A}}^{2}b+2A\ddot{A}b-3A\ddot{A}a)/Aa(A\ddot{A}+2{\dot{A}}^{2})}dt,\\
C(t)&=&\frac{(9A\ddot{A}a^{2}b-5A\ddot{A}a^{3}-4A\ddot{A}b^{2}a+2a^{3}{\dot{A}}^{2}-2b^{2}a{\dot{A}}^{2})}
{Q},\\
T(t)&=&0,\label{a2}
\end{eqnarray}
where
\begin{eqnarray*}
Q=&&40B\dot{A}\ddot{A}Aba+2B{\dot{A}}^{3}ba+2B{\dot{A}}^{3}bam+40B\dot{A}\ddot{A}Abam
-16B\dot{A}\ddot{A}Ab^{2}m-8B{\dot{A}}^{3}b^{2}m\\
&&-8B{\dot{A}}^{3}b^{2}-16B\dot{A}\ddot{A}Ab^{2}-28B\dot{A}\ddot{A}A{a}^{2}m-28B\dot{A}\ddot{A}A{a}^{2}
+4B{\dot{A}}^{3}{a}^{2}m+4B{\dot{A}}^{3}{a}^{2}.
\end{eqnarray*}
and $c_{1}$, $c_{2}$ and $c_{3}$ are arbitrary constants.

\section{Conclusion}

In this paper, we investigated the Noether gauge symmetries of $f(T)$ in a
homogeneous but anisotropic Bianchi type I backgrounds. We solved the
gauge equations and classified the model according to generators. We showed that
there are five symmetries, in which $\textbf{X}_{1}$ corresponds to time invariance of the model,
 $\textbf{X}_{2,3}$ define the rescaling translational invariance and results the conserve momentum,
 $\textbf{X}_{4}$ is a combination of rescaled translational symmetry and finally, $\textbf{X}_{5}$
  shows the contraction in time of the rescaled invariance of the whole model. The related invariances
  $\mathbf{I}_{1}$ is the invariant under time translation, i.e. the energy or Hamiltonian of the system,
   $\mathbf{I}_{2,3}$ characterize some conserve rotational momentums of the system, $\mathbf{I}_{4}$ is
   the rescaling momentum invariance and $\mathbf{I}_{5}$ is the generalized conserved but non-holonomic
   momentum of the system under symmetry $\textbf{X}_{5}$.

\end{document}